 \documentclass[preprint2]{aastex} 

\slugcomment{}

\shorttitle{Does a low solar cycle minimum hint at a weak upcoming
cycle?}
  \shortauthors{Z. L. Du \& H. N. Wang}

\begin{document}
\title{Does a low solar cycle minimum hint at a weak upcoming cycle?}

\author{Z. L. Du and H. N. Wang\altaffilmark{}}
\affil{Key Laboratory of Solar Activity, National Astronomical
Observatories, Chinese Academy of Sciences, Beijing 100012, China}
\email{zldu@nao.cas.cn}

\begin{abstract}
The maximum amplitude ($R_{\mathrm{m}}$) of a solar cycle, in the
term of mean sunspot numbers, is well-known to be positively
correlated with the preceding minimum ($R_{\mathrm{min}}$). So far
as the long term trend is concerned, a low level of
$R_{\mathrm{min}}$ tends to be followed by a weak
$R_{\mathrm{m}}$, and vice versa. In this paper, we found that the
evidence is insufficient to infer a very weak Cycle 24 from the
very low $R_{\mathrm{min}}$ in the preceding cycle. This is
concluded by analyzing the correlation in the temporal variations
of parameters for two successive cycles.
\end{abstract}


\section{Introduction}           
\label{sect:intro}

Studying the correlation between the maximum amplitude
($R_{\mathrm{m}}$) of a solar cycle and the preceding minimum
($R_{\mathrm{min}}$) is useful for understanding the long-term
evolution of solar activity. This can provide information about
the activity level of an ensuing cycle. The positive correlation
between $R_{\mathrm{m}}$ and $R_{\mathrm{min}}$ is a well-known
fact~\citep{Hathaway02}. As a natural consequence, the very low
level of solar activity at the present time (around the onset of
Cycle 24) seems to be followed by a very weak
\citep{Svalgaard05,Schatten05}, or even the weakest cycle
\citep{Li09}. However, a lower $R_{\mathrm{min}}$ has not always
been followed by a weaker cycle. For example, a small
$R_{\mathrm{min}}$ precedes the greatest $R_{\mathrm{m}}$ in Cycle
19~\citep{Wang09}. Therefore, what information we can infer from
the preceding minimum is worth re-analyzing carefully. We ask
whether and how past cycles affect the present cycle.

A more accurate prediction of solar activity is an important task
in solar physics and space weather. Knowing in advance the
activity level of an upcoming cycle is helpful in the launching
and operation of spacecrafts. An underestimate of the activity
level for the next cycle may let down our guard. One aim of this
study serves to remind the space flight mission planners that they
still need to remain vigilant to avoid unexpected troubles.

In the present study, we use the 13-month running mean of
Z\"{u}rich relative sunspot
number\footnote{http://www.ngdc.noaa.gov/stp/SOLAR/ftpsunspot\-number.\-html}
from January 1749 to April 2010 to determine the maximum
($R_{\mathrm{m}}$) and the preceding minimum ($R_{\mathrm{min}}$)
of the solar cycle. The correlation between $R_{\mathrm{m}}$ and
$R_{\mathrm{min}}$ for different periods of time is shown in
Section \ref{sec:Correlation}. In Section \ref{sec:Trends}, we
examine the varying trends of $R_{\mathrm{m}}$ and
$R_{\mathrm{min}}$ using a quantity to describe whether a
parameter increases or decreases. Then, in Section
\ref{sec:Running}, we analyze the temporal variation in the
correlation coefficient between $R_{\mathrm{m}}$ and
$R_{\mathrm{min}}$ with a moving time window of five cycles. The
results are briefly discussed and summarized in Section
\ref{sec:Discussions}.

\section{Correlation between $R_{\mathrm{m}}$
   and $R_{\mathrm{min}}$} \label{sec:Correlation}

The parameters of $R_{\rm m}$ and $R_{\rm min}$ since cycle $n= 1$
are listed in Table~\ref{Table:1}, and shown in Fig.\,1a.

   \begin{figure}[h!!!]
   \centering
   \includegraphics[width=8.0cm, angle=0]{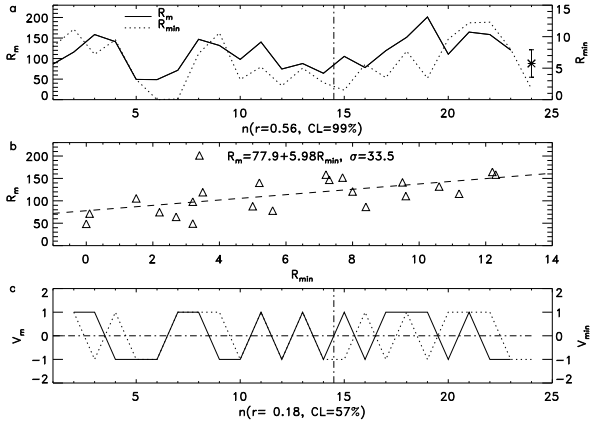}
   \begin{minipage}[]{85mm}
   \caption{ (a)  $R_{\mathrm{m}}$(solid) and $R_{\mathrm{min}}$(dotted) since cycle $n=1$.
     (b) Scatter plot of $R_{\mathrm{m}}$ against $R_{\mathrm{min}}$
     (triangles).
     (c) Varying trend: $V_{\mathrm{m}}$(solid) and $V_{\mathrm{min}}$(dotted). }
      \end{minipage}
   \label{Fig1}
   \end{figure}

It can be seen from Fig.\,1a that $R_{\mathrm{m}}$ and
$R_{\mathrm{min}}$ have a similar long-term variation behavior: a
lower (higher) level of $R_{\mathrm{min}}$ tends to be followed by
a weaker (stronger) $R_{\mathrm{m}}$. Their correlation
coefficient is $r=0.56$ at a confidence level (CL) of 99\%. The
scatter plot of $R_{\mathrm{m}}$ against $R_{\mathrm{min}}$
(triangles) is shown in Fig.\,1b. Their least-squares-fit linear
regression equation is:
\begin{equation}
  \label{Eq:regression}
   R_{\mathrm{m}} = 77.9 + 5.98 R_{\mathrm{min}},\quad
   \sigma=33.5,  
\end{equation}
where $\sigma$ is the standard deviation of the equation.
Substituting the present value of $R_{\mathrm{min}}$ (1.7) into
this equation, the peak of Cycle 24 is predicted as
$R_{\mathrm{m}}(24)= 88.0\pm 33.5$ (labeled by a star), regarding
the 1-$\sigma$ uncertainty. When using the modern era data since
Cycle 10, the peak of Cycle 24 is predicted to be higher, as
$R_{\mathrm{m}}(24)= 98.4\pm 35.0$. When using only the recent
nine cycles since Cycle 15, an even higher value of
$R_{\mathrm{m}}(24)= 122.3\pm 36.5$ will be predicted for Cycle
24.

However, \citet{Li09} inferred a rather low level for Cycle 24,
$R_{\mathrm{m}}(24)= 58.0\pm 26.6$, from a relationship of
$R_{\mathrm{m}} = 48.8+ 5.39 R_{\mathrm{min}}\pm26.6$ derived by
\citet{Hathaway02}. The data used to derive this relationship are
those that are smoothed with the 24-month Gaussian filter, rather
than the 'standard' 13-month running mean. So, the present minimum
value ($R_{\mathrm{min}}=1.7$) of the 13-month running mean
sunspot number is inappropriate for use in inferring an
$R_{\mathrm{m}}(24)$ value from this relationship. Besides, the
$R_{\mathrm{min}}$ value in terms of the 24-month Gaussian filter
is unknown within 12 months of the minimum. (Even if this value is
known, the result inferred from the above relationship has a
different meaning.)

We return to Fig.\,1a. If using only the parameters in the earlier
cycles of $n=1$--14 (left to the vertical line in Fig.\,1a), the
correlation coefficient between $R_{\mathrm{m}}$ and
$R_{\mathrm{min}}$ increases to $r(1$-14$)=0.72$ at the 99\% level
of confidence. In contrast, for the recent cycles of $n=15$--23,
the correlation coefficient is only $r($15-$23)=0.23$, which is
statistically insignificant (CL $<50\%)$. Therefore, the positive
correlation between $R_{\mathrm{m}}$ and $R_{\mathrm{min}}$ (0.56)
is mostly contributed by the earlier cycles. The recent cycles,
especially for cycles 15--19, seem to behave differently from the
earlier cycles. It is then necessary to analyze whether the
temporal variation in the correlation affects the future
$R_{\mathrm{m}}$ value.

\section{Trends of variations in $R_{\mathrm{m}}$
   and $R_{\mathrm{min}}$} \label{sec:Trends}

It should be noted in Fig.\,1a that, for an individual cycle, the
increase or decrease of $R_{\mathrm{m}}$ does not always follow
that of $R_{\mathrm{min}}$. For example, $R_{\mathrm{min}}$
decreases while $R_{\mathrm{m}}$ increases from $n=18$ to 19. To
demonstrate the behavior of increasing or decreasing, we define
the varying trends of $R_{\mathrm{m}}$ and $R_{\mathrm{min}}$ as:
\begin{equation}
  \label{Eq:trend}
  \begin{array}{lll}
   V_{\mathrm{m}}(n)   &=&\mathrm{Sgn} (R_{\mathrm{m}}(n)-R_{\mathrm{m}}(n-1)),\\
   V_{\mathrm{min}}(n) &=&\mathrm{Sgn} (R_{\mathrm{min}}(n)-R_{\mathrm{min}}(n-1)),
   \end{array}
\end{equation}
where $y=\mathrm{Sgn}(x)$ is the sign function: $y=1$ if $x>0$,
$y=-1$ if $x<0$ and $y=0$ if $x=0$. $V_{\mathrm{m}}(n)=+1$ refers
to an increase in $R_{\mathrm{m}}(n)>R_{\mathrm{m}}(n-1)$, and
$V_{\mathrm{m}}(n)=-1$ refers to a decrease in
$R_{\mathrm{m}}(n)<R_{\mathrm{m}}(n-1)$, and so on.
$V_{\mathrm{m}}(n)=V_{\mathrm{min}}(n)$ refers to the same trend
of $R_{\mathrm{m}}(n)$ and $R_{\mathrm{min}}(n)$. The values of
$V_{\mathrm{m}}(n)$ and $V_{\mathrm{min}}(n)$ are listed in
Table~\ref{Table:1} and shown in Fig.\,1c. For all cycles
$n=2$--23, there are 13/9 pairs of $V_{\mathrm{m}}(n)$ and
$V_{\mathrm{min}}(n)$ with the same/opposite trends. Their
correlation is very weak, at $r=0.18$, and statistically
insignificant (CL = 57\%).

For the earlier cycles of $n=2$--14, there are 10/3 pairs of
$V_{\mathrm{m}}(n)$ and $V_{\mathrm{min}}(n)$ with the
same/opposite trends, and their correlation coefficient is
$r_V(2$-$14)=0.55$ at the 95\% level of confidence. In contrast,
for the recent cycles of $n=15$--23, there are 3/6 pairs with the
same/opposite trends, and their correlation coefficient becomes
negative, at $r_V(2$-$14)=-0.35$, at the 63\% level of confidence.
If we consider only the case of $V_{\mathrm{min}}(n)=-1$, as for
$n=3$, 5, 6, 10, 12, 14, 15, 17, 19, and 23, there are 6/4 pairs
with the same/opposite trends. For the earlier cycles of
$n\leq14$, there are 5/1 pairs for $V_{\mathrm{min}}(n)=-1$ with
the same/opposite trends, while for the recent cycles of
$n\geq15$, there are 1/3 pairs for $V_{\mathrm{min}}(n)=-1$ with
the same/opposite trends. This implies that a decrease in
$R_{\mathrm{min}}$ is indeed followed by a decrease in
$R_{\mathrm{m}}$ in most of the earlier cycles, while in the
recent cycles, a decrease of $R_{\mathrm{min}}$ tends to be
followed by an increase of $R_{\mathrm{m}}$.

In summary, in terms of the varying trend, there is no
statistically significant positive correlation between
$V_{\mathrm{m}}$ and $V_{\mathrm{min}}$ for all cycles ($r=0.18$).
The positive correlation between $V_{\mathrm{m}}$ and
$V_{\mathrm{min}}$ exists only in the earlier cycles
($r_V(2$-$14)=0.55$) at the 95\% level of confidence. Concerning
in the recent cycles, this correlation becomes negative
($r_V(2$-$14)=-0.35$). The behavior of the varying trend changed
in recent cycles. Therefore, we cannot infer a decrease of
$R_{\mathrm{m}}$ from a decrease of $R_{\mathrm{min}}$ for Cycle
24.

\section{Temporal Variation in the Running Correlation }
\label{sec:Running}

In the previous two sections, one has noted that the correlation
between $R_{\mathrm{m}}$ and $R_{\mathrm{min}}$ behaves
differently for different periods of time. Now, we analyze the
temporal variation in the running correlation with a moving time
window of $w=5$ cycles. For each cycle $n$, we calculate the
correlation coefficient between $R_{\mathrm{m}}(i)$ and
$R_{\mathrm{min}}(i)$ for $i=n-2, n-1,\ldots,n+2$~\citep{Du309a},
denoted by $r(5,n)$. The results are shown in Fig.\,2a.

   \begin{figure}[h!!!]
   \centering
   \includegraphics[width=7.0cm, angle=0]{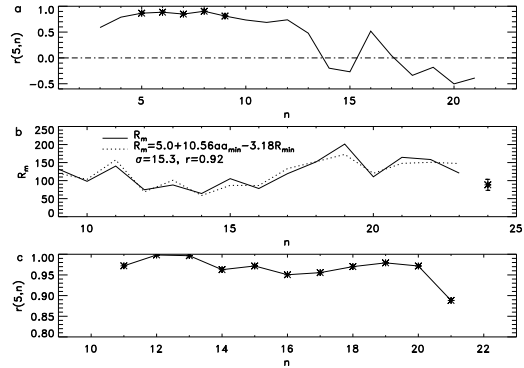}
   \begin{minipage}[]{85mm}
   \caption{ (a) Running correlation coefficient
   between $R_{\mathrm{m}}$ and $R_{\mathrm{min}}$
   for a 5-cycle moving window.
   Stars indicate that the relevant values are significant at the 90\% level of confidence.
   (b) $R_{\mathrm{m}}$ (solid) versa the fitted value (dotted) from
   Equation (\ref{Eq:biv}).
   (c) Running correlation coefficient
   of $R_{\mathrm{m}}$ with both $aa_{\mathrm{min}}$ and $R_{\mathrm{min}}$
   for a 5-cycle moving window.
   }
      \end{minipage}
   \label{Fig2}
   \end{figure}

It can be seen that the correlation is positive before $n=13$ and
significant at the 90\% level of confidence for cycles $n=5$--9
(stars). This implies that a lower (higher) level of
$R_{\mathrm{min}}$ tends to be followed by a weaker (stronger)
$R_{\mathrm{m}}$ for these earlier cycles. However, the
correlation decreases since $n=14$, and becomes negative since
$n=18$, implying that a lower $R_{\mathrm{min}}$ corresponds to a
stronger $R_{\mathrm{m}}$ (see also $n=15$, 17 and 19 in
Fig.\,1a). Therefore, a lower $R_{\mathrm{min}}$ has not always
been followed by a weaker $R_{\mathrm{m}}$. In other words, we
cannot infer a very weak $R_{\mathrm{m}}$ of Cycle 24 from the
preceding very low level of $R_{\mathrm{min}}$.

\section{Discussions and Conclusions}
\label{sec:Discussions}

It is well known that the maximum amplitude ($R_{\mathrm{m}}$) of
a solar cycle is positively correlated with the preceding minimum
($R_{\mathrm{min}}$), so that a low $R_{\mathrm{min}}$ tends to be
followed by a weak $R_{\mathrm{m}}$. However, this relationship is
not always effective for individual cycles~\citep{Wang09},
especially for the recent cycles, as shown in Fig.\,2a. The
correlation between $R_{\mathrm{m}}$ and $R_{\mathrm{min}}$ varies
with time ($n$).

We analyzed the temporal behavior of this correlation and the
varying trends ($V$) of $R_{\mathrm{m}}$ and $R_{\mathrm{min}}$.
In the recent cycles, they all show a negative correlation. Since
the prediction of $R_{\mathrm{m}}$ relies more on the recent cycle
rather than on the  past cycles
\citep{Schatten05,Svalgaard05,Du308,Du309b}, the negative
correlation in the recent cycles cannot infer a very weak
$R_{\mathrm{m}}$ from a very low $R_{\mathrm{min}}$.

One may argue that $R_{\mathrm{m}}$ and $R_{\mathrm{min}}$ have a
similar shape in the most recent four cycles of $n=20$--23. Along
with the developing trend of these cycles, $R_{\mathrm{m}}(24)$
should be very small. However, whether this behavior holds true is
questionable before and after these cycles. It should be noted in
Fig.\,1a that $R_{\mathrm{m}}$ has never decreased in three
successive cycles. The $R_{\mathrm{m}}$ value decreased two cycles
from $n=3$ to 5, and then leveled off to $n=6$, and decreased two
cycles from $n=8$ to 10, and then increased to $n=11$. Now that
the $R_{\mathrm{m}}$ value decreased two cycles from $n=21$ to 23,
it seems to increase or level off according to its past behavior.
On the other hand, $R_{\mathrm{min}}(24)$ is not the lowest one
ever seen. It is higher than cycles 6, 7 (Fig.\,1a), and 15
\citep{Li09}:
$R_{\mathrm{min}}(24)>R_{\mathrm{min}}(15)>R_{\mathrm{min}}(7)>R_{\mathrm{min}}(6)$.
However, corresponding to these local minima, the following
$R_{\mathrm{m}}$ values are not local minima:
$R_{\mathrm{m}}(6)\sim R_{\mathrm{m}}(5)$,
$R_{\mathrm{m}}(7)>R_{\mathrm{m}}(6)$, and
$R_{\mathrm{m}}(15)>R_{\mathrm{m}}(14)$. From this information, we
cannot yet infer that Cycle 24 is a local minimum. To say the
least, it is unlikely that Cycle 24 will be the weakest cycle.

In conclusion, we have not found sufficient evidence for the
low(est) level of Solar Cycle 24 inferred from the low level of
the present state. The sunspot number is highly correlated with
other solar activity indices, such as sunspot group number,
sunspot area, solar radio flux, and so on. Therefore, the above
conclusions can also be reached when using these indices.

Near the time of the solar cycle minimum, geomagnetic activity  is
a much better indicator of the ensuing maximum amplitude
($R_{\mathrm{m}}$) for the sunspot cycle~\citep{Ohl66} than the
minimum amplitude is ($R_{\mathrm{min}}$). \citeauthor{Hathaway99}
(\citeyear{Hathaway99,Hathaway09}) tested the predictive powers of
several methods for cycles 19-23, and concluded that the
geomagnetic-related precursor methods outperform the others.  The
minimum smoothed monthly mean aa index ($aa_{\mathrm{min}}$) near
the time of the solar cycle minimum is shown in Table~1, in which
the values of cycles 9-11 are taken from the equivalent annual
values~\citep{Du309b}. One can note that the varying trend (V) of
$R_{\mathrm{m}}$ follows well with that of $aa_{\mathrm{min}}$ ---
with only the two exceptions of cycles 16 and 22. The correlation
coefficient between $R_{\mathrm{m}}$ and $aa_{\mathrm{min}}$ is
usually as high as 0.9 \citep{Du309b}. The application of
$aa_{\mathrm{min}}$ in the prediction of $R_{\mathrm{m}}$ can be
found, for example, in \citet{Hathaway09} and \citet{Du309b}.
\citet{Wilson98} suggested the bivariate of both
$aa_{\mathrm{min}}$ and $R_{\mathrm{min}}$ to predict
$R_{\mathrm{m}}$. Using the data for cycles 9-23 in Table~1, the
bivariate-fit  regression equation of $R_{\mathrm{m}}$ versus both
$aa_{\mathrm{min}}$ and $R_{\mathrm{m}}$ is:
\begin{equation}
  \label{Eq:biv}
   R_{\mathrm{m}} = 5.0 + 10.56 aa_{\mathrm{min}} -3.18 R_{\mathrm{min}},\quad
   \sigma=15.3,
\end{equation}
where $\sigma$ is the standard deviation of the equation. Figure
~2b shows the observed  $R_{\mathrm{m}}$ (solid) and the fitted
$R_{\mathrm{m}}$ (dotted) from the above equation. Substituting
the values of $aa_{\mathrm{min}}(8.4)$ and $R_{\mathrm{min}}$
(1.7) into this equation, the peak of the next cycle is predicted
as $R_{\mathrm{m}}(24)= 88.3\pm 15.3$ (labeled by a star). This
prediction is close to that predicted by the single variate of
$R_{\mathrm{min}}$ in Equation~(\ref{Eq:regression}). But, the
correlation for the bivariate of both $aa_{\mathrm{min}}$ and
$R_{\mathrm{m}}(r=0.92)$ is much higher than that of the single
variate of $R_{\mathrm{min}}(r=0.56)$. If this prediction comes
true, Cycle 24 will be modest rather than the lowest one.

The prediction of $R_{\mathrm{m}}$ is related to the behavior of
solar activity in the past cycles.  \citet{Du309b}  pointed out
that Ohl's precursor method performed well only if the related
correlation coefficient becomes stronger. If the correlation
coefficient becomes weaker, its prediction would be questionable.
Figure~2c shows the running correlation coefficient $r(5,n)$ of
$R_{\mathrm{m}}$ with both $aa_{\mathrm{min}}$ and
$R_{\mathrm{min}}$   for a five-cycle moving window. It is seen
that the last value ($n=21$, corresponding to the data for cycles
19-23) drops drastically. Therefore, other methods are needed to
check the above prediction.

Predicting the future level of a solar cycle is a complex project
in solar physics and space weather~\citep{Wang09b}. This paper
stresses that the low level of $R_{\mathrm{min}}$ in the present
state is insufficient to infer a low(est) level for Solar Cycle
24, as suggested by \citet{Li09}. Whether a prediction from a
simple parameter succeeds is related to the behavior of solar
activity in the past few cycles. When a solar cycle is well
underway (two to three years after the minimum), its behavior can
be predicted to a good extent with curve fitting
techniques~\citep{Hathaway09}.

\section*{Acknowledgments}
The authors are grateful to an anonymous referee for useful
comments. This work is supported by Chinese Academy of Sciences
through grant KJCX2\,--\,YWT04, and the National Natural Science
Foundation of China through grants 10973020  and 40890161.



\begin{table}[h!!!]
\small \centering
\begin{minipage}[]{120mm}
  \caption[]{ Cycle maximum ($R_{\rm m}$), minimum ($R_{\rm min}$) and minimum aa index ($aa_{\rm min}$) and their trends (V). 
 }\label{Table:1}\end{minipage}
\tabcolsep 4.5mm  
 \begin{tabular}{rrrr|rrrr}
  \hline\noalign{\smallskip}
$n$ & $R_{\rm m}(V)$ & $R_{\rm min}(V)$ &  $aa_{\rm min}$ &  $n$ & $R_{\rm m}(V)$ &  $R_{\rm min}(V)$ & $aa_{\rm min}$  \\
  \hline\noalign{\smallskip}
 1  & {86.5\quad\ \ \,} & {8.4\quad\ \ \,} &                     & 13 &  $87.9(+)$  &  $5.0(+)$ & $10.6(+)$  \\
 2  & $115.8(+)$        & $11.2(+)$        &                     & 14 &  $64.2(-)$  &  $2.7(-)$ & $5.9(-)$ \\
 3  & $158.5(+)$        &  $7.2(-)$        &                     & 15 & $105.4(+)$  &  $1.5(-)$ & $8.2(+)$ \\
 4  & $141.2(-)$        &  $9.5(+)$        &                     & 16 &  $78.1(-)$  &  $5.6(+)$ & $9.4(+)$ \\
 5  &  $49.2(-)$        &  $3.2(-)$        &                     & 17 & $119.2(+)$  &  $3.5(-)$ & $13.2(+)$ \\
 6  &  $48.7(-)$        &  $0.0(-)$        &                     & 18 & $151.8(+)$  &  $7.7(+)$ & $16.3(+)$ \\
 7  &  $71.5(+)$        &  $0.1(+)$        &                     & 19 & $201.3(+)$  &  $3.4(-)$ & $16.9(+)$ \\
 8  & $146.9(+)$        &  $7.3(+)$        &                     & 20 & $110.6(-)$  &  $9.6(+)$ & $13.8(-)$ \\
 9  & $131.9(-)$        & $10.6(+)$        & {14.1\quad\ \ \,}   & 21 & $164.5(+)$  & $12.2(+)$ & $17.2(+)$ \\
10  & $98.0(-)$         &  $3.2(-)$        & $10.3(-)$           & 22 & $158.5(-)$  & $12.3(+)$ & $17.5(+)$ \\
11  & $140.3(+)$        &  $5.2(+)$        & $16.0(+)$           & 23 & $120.8(-)$  &  $8.0(-)$ & $15.9(-)$ \\
12  &  $74.6(-)$        &  $2.2(-)$        & $6.7(-)$            & 24 &   ?\quad(?) &  $1.7(-)$ & $8.4(-)$   \\
  \noalign{\smallskip}\hline
\end{tabular}
\end{table}

\end{document}